\documentclass[aps,prd,floats,epsf,superscriptaddress,nofootinbib,amsmath]{revtex4}
\usepackage{amsmath,amssymb}
\usepackage{graphicx}
\usepackage{multirow}
\usepackage{slashed}

\begin{document}

{\small
\begin{flushright}
IPMU15-0079\\
\end{flushright} }

\bigskip

\title{Standard Model Effective Field Theory: \\
Integrating out Vector-Like Fermions}

\author{Ran Huo}
\email[e-mail: ]{ran.huo@ipmu.jp}
\affiliation{Kavli IPMU (WPI), UTIAS, The University of Tokyo, Kashiwa, Chiba 277-8583, Japan}

\date{\today}

\begin{abstract}
We apply the covariant derivative expansion of the Coleman-Weinberg potential to vector-like fermion models, matching the UV theory to the relevant dimension-6 operators in the standard model effective field theory. The $\gamma$ matrix induced complication in the fermionic covariant derivative expansion is studied in detail, and all the contributing combinations are enumerated. From this analytical result we also provide numerical constraints for a generation of vector-like quarks and vector-like leptons.
\end{abstract}

\maketitle

\section{Introduction}

The discovery of the 125-GeV Higgs boson with properties consistent with standard model (SM) predictions eventually proves the validity of the SM. New physics should reside at a sufficiently high mass scale and/or interact with the SM particles through very small couplings, so that it does not affect the SM sector significantly. Thus an effective field theory (EFT) of the SM, with all fields appearing in the operators being only the SM ones and all the new physics integrated out, would be an ideal tool to study the low energy physics. Today it is well known that going beyond the only dimension-5 operator in the seesaw mechanism, up to dimension-6 level such an EFT consists a total of 59 independent operators~\cite{Grzadkowski:2010es} for one family of fermions as a complete basis, while five more can be added if baryon number violation is allowed.

In the SM EFT collider or other experiment observables can be calculated to the lowest order of the dimension-6 operators~\cite{Burgess:1993vc}, and a global fit~\cite{Han:2004az} will determine the Wilson coefficients of the operators in a model independent way. On the other hand, if one is interested in a specific model, then translating these model independent constraints to model parameters calls for another kind of matching. The most famous example is the Peskin-Takeuchi oblique corrections of $S$ and $T$ parameters~\cite{Peskin:1990zt}\footnote{The $U$ parameter defined there corresponds to dimension-8 operator.}, and the $T$ parameter can even date back to the context of weak charged current and neutral current ratio in the form of $\rho$ parameter. Another example is the Higgs diphoton decay branching ratio calculations in the early days of Higgs discovery, hinted by the ATLAS and CMS measured Higgs diphoton channel signal strength. In the following we will see how they are equivalent to the Wilson coefficients of the dimension-6 operators.

Usually the calculations are done case by case, using the traditional Feynman diagram approach. To the best knowledge of the author, the first ``systematic'' dimension-6 operator calculations are given in~\cite{Henning:2014gca,Henning:2014wua}. Instead of the Feynman diagram technique, they use the so called covariant~\cite{Gaillard:1985uh,Cheyette:1987qz} derivative expansion~\cite{Iliopoulos:1974ur,Novikov:1983gd,Fraser:1984zb,Zuk:1985sw,Chan:1985ny,Cheyette:1985ue} (CDE) of the Coleman-Weinberg (CW) potential~\cite{Coleman:1973jx}, which has the merit of generating all the dimension-6 SM operators automatically at one-loop level, including the oblique $S$ and $T$, the Higgs relevant ones and all the other ones involving the Higgs and SM gauge bosons. In this paper we will apply the same technique to a prototype of fermionic model\footnote{For scalar models, further see~\cite{Chiang:2015ura}.}, namely a mirror vector-like (VL) fermion sector. By definition the mirror VL fermion should contain both doublet and two singlets of $SU(2)_L$, but we consider them to be in arbitrary representation (denoted as $m$, with $m=1,3,8,6$ \emph{etc}) of the $SU(3)_c$ and the $SU(2)_L$ doublet to carry arbitrary $U(1)_Y$ hypercharge $Y$.

Guided by appendix of~\cite{Henning:2014wua}, the fermionic CDE has extra $\gamma$ matrix related contributions. Trace over the $\gamma$ matrix space immediately picks out combinations of only even number of $\gamma$ matrices, leaving odd number ones vanishing. It is like another ``power counting'' in addition to the one of dimension of the operators, and the bosonic CDE corresponds only to the case of no $\gamma$ matrix. We will see that up to dimension-6 operators there are only two $\gamma$ and four $\gamma$ matrices contribution in extra.

This paper is organized as follows. In Section~\ref{sec:for}, we list the dimension-6 operators and the bosonic CDE formula, then we generalize it to the fermionic case by working out all possible extra $\gamma$ matrix related terms. Then in Section~\ref{sec:res} we define the VL fermion model and show the main analytical results, as well as some numerical results for the most interesting VL lepton and quark model. We discuss the formulism and conclude in Section~\ref{sec:dis}. At last, Appendix~\ref{sec:col} provides a complete list of contributing combinations, operator by operator in the dimension-6 operator basis and term by term in the CDE, in order to facilitate future calculation.

\section{Formulism\label{sec:for}}

\begin{table}[th]
\begin{tabular}{cccccccc}
\hline\hline
~Symbol~ & Operator expression &~~& ~Symbol~ & Operator expression &~~& ~Symbol~ & Operator expression\\
\hline
${\cal O}_6$ & $(H^\dag H)^3$ && ${\cal O}_{GG}$ & $g_s^2H^\dag HG^a_{\mu\nu}G^{a\mu\nu}$ && ${\cal O}_{W}$ & $ig(H^\dag\overleftrightarrow{D}_\mu t^a H )D_\nu W^{a\mu\nu}$ \\
${\cal O}_H$ & $\frac12(\partial_\mu (H^\dag H))^2$ && ${\cal O}_{WW}$ & $g^2H^\dag HW^a_{\mu\nu}W^{a\mu\nu}$ && ${\cal O}_{B}$ & $ig'(H^\dag\overleftrightarrow{D}_\mu H)\partial_\nu B^{\mu\nu}$ \\
${\cal O}_T$ & $\frac12(H^\dag \overleftrightarrow{D}_\mu H)^2$ && ${\cal O}_{BB}$ & $g'^2H^\dag HB_{\mu\nu}B^{\mu\nu}$ && ${\cal O}_{HW}$ & $2ig(D_\mu H)^\dag t^a(D_\nu H)W^{a\mu\nu}$ \\
${\cal O}_R$ & $(H^\dag H)(D_\mu H^\dag D^\mu H)$ && ${\cal O}_{WB}$ & $2gg'H^\dag t^a HW_{\mu\nu}^a B^{\mu\nu}$ && ${\cal O}_{HB}$ & $2ig'Y_H(D_\mu H)^\dag(D_\nu H)B^{\mu\nu}$ \\
 && & && & ${\cal O}_D$ & $(D_\mu D^\mu H^\dag)(D_\nu D^\nu H)$\\
\hline\hline
\end{tabular}
\caption{Independent CP-even dimension-6 operators composed of only the Higgs and gauge boson fields that are relevant to the analysis in this work. Notations of fields and operators are explained in the main text.
\label{tab:Op}}
\end{table}
The operator basis we use is listed in Table~\ref{tab:Op}, which includes only Higgs and electroweak gauge boson fields. For operators involving SM fermions generally we don't need the following technique and a tree level matching~\cite{delAguila:2000rc} will suffice. Apparently this is a redundant basis, for example we have
\begin{equation}
\frac{1}{4}(\mathcal{O}_{WW}+\mathcal{O}_{WB})-\mathcal{O}_{W}+\mathcal{O}_{HW}=0~,\qquad\frac{1}{4}(\mathcal{O}_{BB}+\mathcal{O}_{WB})-\mathcal{O}_{B}+\mathcal{O}_{HB}=0~,
\label{eq:HWB}
\end{equation}
so that the operator $\mathcal{O}_{HW}$ and $\mathcal{O}_{HB}$ as used in~\cite{Hagiwara:1993ck}\footnote{Since the oblique $S$ parameter is given by the dimension-6 Wilson coefficients of $S=4\pi v^2(4c_{WB}+c_W+c_B)$, we can see that $\mathcal{O}_{HW}$ and $\mathcal{O}_{HB}$ do not contribute to the S parameter. Similarly they will not contribute to the Higgs diphoton branching ratio.} can be switched into $\mathcal{O}_W$, $\mathcal{O}_B$ as well as $\mathcal{O}_{WW}$, $\mathcal{O}_{BB}$ and $\mathcal{O}_{WB}$. However, the complete list in Table~\ref{tab:Op} has an advantage of directly providing a one to one correspondence to all dimension-6 operator generated in the CDE calculation.

Recall that for the bosonic CW potential
\begin{equation}
V_\text{CW}=-\frac{in_B}{2}\int\frac{d^dp}{(2\pi)^d}\ln(p^2-V''),\label{eq:CW}
\end{equation}
the CDE~\cite{Henning:2014wua} can be written as a textbook level loop integration
\begin{equation}
\mathcal{L}_{\text{CDE,}B}=\frac{n_B}{2}\int_0^\infty du\int\frac{d^dp_E}{(2\pi)^d}\sum_{m=1}^\infty(-1)^m\text{tr}\bigg[\bigg(\frac{1}{p_E^2+M^2+u}\Big[\delta\tilde{V}''+\tilde{G}\Big]\bigg)^m\frac{1}{p_E^2+M^2+u}\bigg].\label{eq:CDEB}
\end{equation}
Here $n_B$ is the bosonic degree of freedom (DOF) as in the CW potential, $1$ for a real bosonic DOF and $2$ for a complex bosonic DOF. We have assumed dimensional regularization in the loop integration, and in the CDE a Wick rotation is performed and subscript ``E'' indicates Euclidean. $V$ and $G$ generally denote potential and gauge terms respectively, and the double prime on $V$ indicates two derivatives with any beyond standard model (BSM) fields. The CDE assumes $V''=M^2+\delta V''$ and $M^2$ is some large constant squared mass term which is irrelevant to Higgs vacuum expectation value (VEV), $\delta V''$ on the other hand picks the spacetime dependent part of potential terms. Here not only the Higgs VEV but both charged and neutral Higgs components are counted. The SM Higgs sector as well as the gauge field can couple between different components of BSM fields, and the $\delta V''$ should be generalized into matrix, with each entry corresponding to coupling term quadratic in BSM component fields. So a trace over the matrix basis is needed. 
$\tilde{\enspace}$ generally indicates a Baker-Campbell-Hausdorff (BCH) expansion with covariant derivatives and $\frac{\partial}{\partial p}$s, for example\footnote{In $\tilde{G}$ we only show terms which can contribute to our dimension-6 operators in Table~\ref{tab:Op}, so are the followings.}
\begin{align}
\delta\tilde{V}''=&e^{-iD\frac{\partial}{\partial p}}\delta V''e^{iD\frac{\partial}{\partial p}}=\delta V''+\sum_{n=1}^\infty\frac{(-i)^n}{n!}\thinspace D_{\mu_1}\cdots D_{\mu_n}\delta V''\enspace{\textstyle\frac{\partial}{\partial p}^{\mu_1}\cdots\frac{\partial}{\partial p}^{\mu_n}},\label{eq:expV}\\
\tilde{G}=&gp^\mu t^a\bigg(iF_{\nu\mu}^a{\textstyle\frac{\partial}{\partial p}^\nu}+\frac{4}{3!}D_\rho F_{\nu\mu}^a{\textstyle\frac{\partial}{\partial p}^\rho\frac{\partial}{\partial p}^\nu}+\cdots\bigg)+gt^a\bigg(\frac{2}{3!}D^\mu F_{\nu\mu}^a{\textstyle\frac{\partial}{\partial p}^\nu}+\cdots\bigg)+g^2t^at^b\bigg(\frac{1}{4}F^a_{\nu\mu}F^{b\rho\mu}{\textstyle\frac{\partial}{\partial p}^\nu\frac{\partial}{\partial p}_\rho}+\cdots\bigg).\label{eq:expG}
\end{align}
Here $D$ generally denote a covariant derivative, taking the corresponding gauge field when acting on specific field. The generally non-abelian field strength appears as commutator of covariant derivatives of $F_{\nu\mu}^at^a=\frac{i}{g}[D_\nu,\thinspace D_\mu]$, and the Abelian case is easily got. At last $\mu$ is a dimension-2 auxiliary number, served for the regularization of the order of the expansion.

In the fermionic case Eq.~(\ref{eq:CW}) becomes
\begin{equation}
V_\text{CW}\propto\frac{i}{2}\int\frac{d^dp}{(2\pi)^d}\ln(\slashed{p}-V''_F).\label{eq:CWF}
\end{equation}
On the other hand, the same contribution comes from replacing $\slashed{p}-V''_F$ by $-\slashed{p}-V''_F$. Summing the two logarithms and ignoring the $-1$ inside the logarithm, one gets back to Eq.~(\ref{eq:CW}). Namely the CW potential uniformly describes both bosonic and fermionic models.

We are generalizing the CDE to Eq.~(\ref{eq:CWF}). For the eligibility of the DE one also needs a large spacetime independent term, so similarly $V''_F=M+\delta V''_F$ and $M$ is the VL fermion mass. The spacetime dependent $\delta V''_F$ can only be linear in Higgs, up to a dimensionless coupling. In the $\ln(\slashed{p}-M-\delta V''_F)+\ln(-\slashed{p}-M-\delta V''_F)=\ln((\slashed{p}-M-\delta V''_F)(\slashed{p}+M+\delta V''_F))+\cdots$, there is always a mixture of dimension-2 and dimension-1 terms which corresponds to the bosonic $\delta V''$
\begin{equation}
\delta V''=\{M,\delta V''_F\}+(\delta V''_F)^2~.\label{eq:v''F}
\end{equation}
In this paper $\{,\}$ and $[,]$ denote anticommutator and commutator respectively. The BCH expansion is obtained similarly.

However, as initiated in the appendix of~\cite{Henning:2014wua}, the presence of $\gamma$ matrices introduces further complication, which will become our focus from now on. Recall that the CDE always acts on the canonical momentum of $p_\mu+iD_\mu$ rather than on the kinetic momentum, giving
\begin{align}
e^{-iD\frac{\partial}{\partial p}}\big(p_\mu+iD_\mu\big)e^{iD\frac{\partial}{\partial p}}&=\sum_{n=0}^\infty\frac{1}{n!}[-iD{\textstyle\frac{\partial}{\partial p}},[-iD{\textstyle\frac{\partial}{\partial p}},[\cdots[-iD{\textstyle\frac{\partial}{\partial p}},p_\mu+iD_\mu]\cdots]]]\nonumber\\
&=p_\mu+g\sum_{n=0}^\infty\frac{(-i)^{n+1}(n+1)}{(n+2)!}\thinspace D_{\mu_1}D_{\mu_2}\cdots D_{\mu_n}F_{\nu\mu}\enspace{\textstyle\frac{\partial}{\partial p}^{\mu_1}\cdots\frac{\partial}{\partial p}^{\mu_n}\frac{\partial}{\partial p}^\nu}\nonumber\\
&=p_\mu-\frac{i}{2}gF_{\nu\mu}{\textstyle\frac{\partial}{\partial p}^\nu}-\frac{2}{3!}gD_\rho F_{\nu\mu}{\textstyle\frac{\partial}{\partial p}^\rho\frac{\partial}{\partial p}^\nu}+\frac{3i}{4!}gD_\sigma D_\rho F_{\nu\mu}{\textstyle\frac{\partial}{\partial p}^\sigma\frac{\partial}{\partial p}^\rho\frac{\partial}{\partial p}^\nu}+\cdots.
\label{eq:expP}
\end{align}
The BCH expansion of the first $p_\mu$ term is acted by the $D\frac{\partial}{\partial p}$ and is always less in multiplicity of the commutators by one than the corresponding action on the second $iD_\mu$ term, so comes the $(-i)^{n+1}(n+1)/(n+2)!$ factor. The square of the expression gives the $-\tilde{G}$ expression defined in Eq.~(\ref{eq:expG}), where the first kinetic momentum square term should not be counted. On the other hand, the CDE acting on the potential terms is given by the same expression as Eq.~(\ref{eq:expV}).
\begin{equation}
e^{-iD\frac{\partial}{\partial p}}\delta V''_Fe^{iD\frac{\partial}{\partial p}}=\delta V''_F+\sum_{n=1}^\infty\frac{(-i)^n}{n!}\thinspace D_{\mu_1}\cdots D_{\mu_n}\delta V''_F\enspace{\textstyle\frac{\partial}{\partial p}^{\mu_1}\cdots\frac{\partial}{\partial p}^{\mu_n}}~.\label{eq:expVF}\\
\end{equation}
Then we only need to use the two equations to calculate
\begin{equation}
e^{-iD\frac{\partial}{\partial p}}\bigg((\slashed{p}+i\slashed{D})(\slashed{p}+i\slashed{D})+(\slashed{p}+i\slashed{D})(M+\delta V_F'')-(M+\delta V_F'')(\slashed{p}+i\slashed{D})-(M+\delta V_F'')^2\bigg)e^{iD\frac{\partial}{\partial p}}~.
\end{equation}
All the relevant terms contributing to the fermionic CDE, including the $\gamma$ matrix relevant ones, are generated by applying the BCH expansion.

At first we list the cross term of $\slashed{p}+i\slashed{D}$ with $M+\delta V''_F$, which can be regarded as a commutator $[e^{-iD\frac{\partial}{\partial p}}(\slashed{p}+i\slashed{D})e^{iD\frac{\partial}{\partial p}}, e^{-iD\frac{\partial}{\partial p}}\delta V''_Fe^{iD\frac{\partial}{\partial p}}]$. Expanding the commutator between terms given by of Eq.~(\ref{eq:expVF}) and Eq.~(\ref{eq:expP}) one by one, we get
\begin{align}
\pm\tilde{\Gamma}_1&=i\slashed{D}\delta V''_F+\frac{1}{2}(\slashed{D}D_\mu+D_\mu\slashed{D})\delta V''_F{\textstyle\frac{\partial}{\partial p}^\mu}-\frac{i}{6}(\slashed{D}D_\mu D_\nu+D_\mu\slashed{D}D_\nu+D_\mu D_\nu\slashed{D})\delta V''_F{\textstyle\frac{\partial}{\partial p}^\mu\frac{\partial}{\partial p}^\nu}+\cdots\nonumber\\
&+\frac{i}{2}g\gamma^\mu[\delta V''_F,\thinspace F^a_{\nu\mu}t^a]{\textstyle\frac{\partial}{\partial p}^\nu}+\frac{1}{2}g\gamma^\mu[D_\rho\delta V''_F,\thinspace F^a_{\nu\mu}t^a]{\textstyle\frac{\partial}{\partial p}^\nu\frac{\partial}{\partial p}^\rho}+\frac{2}{3!}g\gamma^\mu[\delta V''_F,\thinspace D_\rho F^a_{\nu\mu}t^a]{\textstyle\frac{\partial}{\partial p}^\nu\frac{\partial}{\partial p}^\rho}+\cdots.
\label{eq:Gamma1}
\end{align}
The overall sign is up to convention. The first line counts only the commutators involving the first $\slashed{p}$ in Eq.~(\ref{eq:expP}), which is proportional to an identity matrix, so that the matrix structure induced commutator vanishes and just the $\frac{\partial}{\partial p}$ action on $\slashed{p}$ remains. Note that generally other terms in the expansion has a matrix structure in the BSM particle basis, especially the matrix structure of $\delta V''_F$ and its derivatives do not commute with the gauge generator $t^a$, so the second line is nonvanishing.

Then we consider the two $\slashed{p}+i\slashed{D}$ terms. We use the identity $\gamma^\mu\gamma^\nu=\frac{1}{2}\{\gamma^\mu,\gamma^\nu\}+\frac{1}{2}[\gamma^\mu,\gamma^\nu]=g^{\mu\nu}+\frac{1}{2}[\gamma^\mu,\gamma^\nu]$, and the contraction with the first $g^{\mu\nu}$ always reproduces the bosonic terms. So in addition we have the extra terms of $\frac{1}{2}[\gamma^\mu,\gamma^\nu]$ contraction with the two $\slashed{p}+i\slashed{D}$ expansions of Eq.~(\ref{eq:expP}), which is
\begin{align}
\tilde{\Gamma}_2&=-\frac{i}{4}g[\gamma^\mu,\gamma^\nu]F_{\mu\nu}^at^a
+\cdots.
\label{eq:Gamma2}
\end{align}
The matrix structure of each term in the expansion is proportional to gauge generator $t^a$ and identical to two terms in the commutator only except the case that the first $\slashed{p}$ is used instead, so the only nonvanishing commutator comes from each $\frac{\partial}{\partial p}$ from the first expansion acting on the $p_\nu$ of the second expansion. Eventually very limited terms contribute in the CDE.

In all, the fermionic counterpart of Eq.~(\ref{eq:CDEB}) reads
\begin{equation}
\mathcal{L}_{\text{CDE,}F}=\frac{n_F}{2}\int_0^\infty du\int\frac{d^dp_E}{(2\pi)^d}\sum_{m=1}^\infty(-1)^m\text{tr}\bigg[\bigg(\frac{1}{p_E^2+M^2+u}\Big[\delta\tilde{V}''+\tilde{G}+\tilde{\Gamma}_1+\tilde{\Gamma}_2\Big]\bigg)^m\frac{1}{p_E^2+M^2+u}\bigg]~.\label{eq:CDEF}
\end{equation}
$n_F=-4$ for a Dirac fermion, and the terms to be expanded are in given in Eq.~(\ref{eq:expV},~\ref{eq:v''F},~\ref{eq:expG},~\ref{eq:Gamma1},~\ref{eq:Gamma2}).

Before finishing this section let us describe the way of evaluating the new $\tilde{\Gamma}_1+\tilde{\Gamma}_2$ terms. When we identify a combination contributing to our target dimension-6 operators, we should first take the trace over the $\gamma$ matrix space. If one takes the usual $4\times4$ matrix representation of the $\gamma$ matrix as in~\cite{Henning:2014wua}, one have $\text{tr}(\gamma^\mu\gamma^\nu)=4g^{\mu\nu}$. And this $4\times4$ matrix representation induces every term of $\delta\tilde{V}''+\tilde{G}$ an internal $4\times4$ identity matrix in the $\gamma$ matrix space, making every term multiplied by a factor of 4. On the other hand, we can take the point of view that the overall normalization of the CW potential is still determined by the traditional way of counting DOF, such that a Dirac fermion counts $-4$. In this way the internal $\gamma$ matrix space is not expanded and we can equivalently take $\text{tr}(\gamma^\mu\gamma^\nu)=g^{\mu\nu}$. Eq.~(\ref{eq:CDEF}) is presented in this form.

\section{Model and Results\label{sec:res}}

Having worked out the general formula, for calculation of a practical model we only need to input the coupling matrix $\delta V''_F$ which is linear in the SM Higgs component fields, and the SM gauge field coupling matrix with the matrix structure basically given by the gauge group generator $t^a$ in corresponding representation. The SM Higgs always couples to chiral fermions, and the BSM particles to be integrated out as coupled to each entry of $\delta V''_F$ should be a Dirac fermion as a requirement of the CW potential, so a chiral projection such as $P_L=\frac{1-\gamma_5}{2}$ and $P_R=\frac{1+\gamma_5}{2}$ is need. In the calculation performed through Eq.~(\ref{eq:CDEF}), one should use the symbolic projection algebra $P_L^2=P_L,P_R^2=P_R,P_LP_R=0,P_L+P_R=1$ repeatedly.

As a prototype we study one generation of mirror VL fermions, which contains Dirac fermions of one $SU(2)_L$ doublet and two singlets. We will neither consider any mixing of the new VL fermions with the SM fermions induced by Yukawa interactions, nor bother with any further UV completion issue such as the vacuum stability and the gauge coupling unification. The most general Lagrangian in the interaction basis is
\begin{eqnarray}
\mathcal{L}&=&\text{Massless pure kinetic terms}-M_D(\bar{u}_Du_D+\bar{d}_Dd_D)-M_{Su}\bar{u}_Su_S-M_{Sd}\bar{d}_Sd_S\nonumber\\
&-&\bigg(\hat{y}_d(u^\dag_{DL},d^\dag_{DL})\Big(\begin{array}{c}H^+ \\ H_0\end{array}\Big)d_{SR}+\hat{y}_u(u^\dag_{DL},d^\dag_{DL})\Big(\begin{array}{c}H_0^\ast \\ -H^- \end{array}\Big)u_{SR}+\check{y}_dd^\dag_{SL}(H^-,H_0^\ast)\Big(\begin{array}{c}u_{DR} \\ d_{DR} \end{array}\Big)\nonumber\\
&&\quad+\check{y}_uu^\dag_{SL}(H_0,-H^+)\Big(\begin{array}{c}u_{DR} \\ d_{DR} \end{array}\Big)+h.c.\bigg)\nonumber\\
&+&{\textstyle\frac{1}{\sqrt{2}}g(\bar{u}_D\slashed{W}^+d_D+\bar{d}_D\slashed{W}^-u_D)}\nonumber\\
&+&{\textstyle\sqrt{g^2+g'^2}\Big(\bar{u}_D(\frac{1}{2}-Q_us_W^2)\slashed{Z}^0u_D+\bar{d}_D(-\frac{1}{2}-Q_ds_W^2)\slashed{Z}^0d_D+\bar{u}_S(-Q_us_W^2)\slashed{Z}^0u_S+\bar{d}_S(-Q_ds_W^2)\slashed{Z}^0d_S\Big)}\nonumber\\
&+&\text{Possible coupling terms with gluon}~,
\end{eqnarray}
where subscript ``D'' and ``S'' indicate the $SU(2)_L$ doublet or singlet, and ``L'' and ``R'' in Yukawa terms indicate the chirality (we changed to Weyl fermion representation for the Yukawa terms). $Q_u=\frac{1}{2}+Y$ and $Q_d=-\frac{1}{2}+Y$ are generic electric charge of the upper and lower components of the weak doublet, with $Y$ the hypercharge of the $SU(2)_L$ doublet. Note that there are two up types and two down types Yukawa couplings due to the VL nature.

With the help of chiral projections from the above Lagrangian we can write the fermionic coupling matrix
\begin{equation}
M+\delta V''_F=\left(\begin{array}{cccc}
M_D & 0 & (\hat{y}_uP_R+\check{y}^\ast_uP_L)H_0^\ast & (\hat{y}_dP_R+\check{y}^\ast_dP_L)H^+\\
0 & M_D & -(\hat{y}_uP_R+\check{y}^\ast_uP_L)H^- & (\hat{y}_dP_R+\check{y}^\ast_dP_L)H_0\\
(\hat{y}^\ast_uP_L+\check{y}_uP_R)H_0 & -(\hat{y}^\ast_uP_L+\check{y}_uP_R)H^+ & M_{Su} & 0 \\
(\hat{y}^\ast_dP_L+\check{y}_dP_R)H^- & (\hat{y}^\ast_dP_L+\check{y}_dP_R)H_0^\ast & 0 & M_{Sd}
\end{array}\right).
\end{equation}
The electroweak gauge field coupling matrix, on the other hand, have the upper left $2$ by $2$ corner filled with $SU(2)_L$ gauge field, and the whole diagonal filled with $U(1)_Y$ gauge field.

We make the simplification of $\hat{y}_u=\check{y}^\ast_u=y_u$, $\hat{y}_d=\check{y}^\ast_d=y_d$ and $M_D=M_{Su}=M_{Sd}=M$. Then the chiral projection is trivial, and the VL mass is proportional to identity matrix, always commuting with other matrices. The CDE of the CW potential takes the most simplified form. Collecting all the contributions as described in the appendix, we get the results
\begin{align}
\mathcal{L}&\supset\frac{m}{(4\pi)^2}\bigg[-\frac{2(|y_u|^6+|y_d|^6)}{15M^2}\mathcal{O}_6\nonumber\\
&\qquad\qquad\enspace-\frac{28(|y_u|^4+|y_d|^4)-12|y_u|^2|y_d|^2}{15M^2}\mathcal{O}_H
+\frac{2(|y_u|^2-|y_d|^2)^2}{5M^2}\mathcal{O}_T
-\frac{34(|y_u|^4+|y_d|^4)+24|y_u|^2|y_d|^2}{15M^2}\mathcal{O}_R\nonumber\\
&\qquad\qquad\enspace-\frac{|y_u|^2+|y_d|^2}{48M^2}\mathcal{O}_{WW}
-\frac{(1+16Y+32Y^2)|y_u|^2+(1-16Y+32Y^2)|y_d|^2}{48M^2}\mathcal{O}_{BB}\nonumber\\
&\qquad\qquad\enspace+\frac{(3+8Y)|y_u|^2+(3-8Y)|y_d|^2}{24M^2}\mathcal{O}_{WB}\nonumber\\
&\qquad\qquad\enspace+\frac{7(|y_u|^2+|y_d|^2)}{60M^2}\mathcal{O}_W+\frac{7(|y_u|^2+|y_d|^2)}{60M^2}\mathcal{O}_B\nonumber\\
&\qquad\qquad\enspace+\frac{53(|y_u|^2+|y_d|^2)}{20M^2}\mathcal{O}_{HW}+\frac{53(|y_u|^2+|y_d|^2)}{20M^2}\mathcal{O}_{HB}\nonumber\\
&\qquad\qquad\enspace+\frac{|y_u|^2+|y_d|^2}{15M^2}\mathcal{O}_D\bigg]\nonumber\\
&+\frac{1}{(4\pi)^2}\frac{f(m)(|y_u|^2+|y_d|^2)}{M^2}\mathcal{O}_{GG}~,
\bigg.
\label{eq:res}
\end{align}
where $m=1,3,8,6$ \emph{etc} is the representation of the $SU(3)_c$, $f(1)=0$, $f(3)=-\frac{1}{3}$, $f(8)=-2$, $f(6)=-\frac{5}{3}$~\footnote{
In~\cite{Henning:2014wua} it is pointed out that there are additional universal contributions to the pure gauge dimension-6 operators ${\cal O}_{2B}$, ${\cal O}_{2W}$, ${\cal O}_{3W}$, ${\cal O}_{2G}$ and ${\cal O}_{3G}$ defined in the reference. The last two will not affect the electroweak and Higgs physics, while the first three are usually small in effect because they are proportional to the SM gauge couplings. For completeness, we also quote the general result here and include them in the following fits:
\begin{equation}
\mathcal{L}\supset-\frac{m}{(4\pi)^2}\bigg[\frac{(2Y^2+(Y+\frac{1}{2})^2+(Y-\frac{1}{2})^2)g'^2}{30M^2}{\cal O}_{2B}+\frac{g^2}{30M^2}{\cal O}_{2W}+\frac{g^2}{30M^2}{\cal O}_{3W}\bigg] ~.
\label{eq:hitoshi}
\end{equation}
}.

Part of the results can be checked against previous ones. BSM physics contribution to the electroweak oblique corrections is related to the dimension-6 operator Wilson coefficient as
\begin{equation}
T=\frac{1}{\alpha_\text{EM}}v^2c_T~,\qquad\qquad S=4\pi v^2(4c_{WB}+c_W+c_B)~,
\label{eq:oblique}
\end{equation}
where $v=246$~GeV is the SM Higgs VEV. The electroweak oblique corrections are calculated in~\cite{Lavoura:1992np,Dawson:2012di} for a VL fermion sector. In the simplification of $\hat{y}_u=\check{y}^\ast_u=y_u$, $\hat{y}_d=\check{y}^\ast_d=y_d$ and $M_D=M_{Su}=M_{Sd}=M$ it is not hard to expand to order of $M^{-2}$ to get analytical formulas, which agrees with the results given by Eq.~(\ref{eq:res}) and (\ref{eq:oblique}). It is pointed out in~\cite{Joglekar:2012vc,ArkaniHamed:2012kq} that the Higgs diphoton decay branching ratio will be enhanced for a VL lepton sector. With the help of Table 8, 9, 10, 11 of~\cite{Henning:2014wua} we can connect the above Wilson coefficients to the Higgs precision measurement observables, and the charged lepton Yukawa correction to the Higgs diphoton decay amplitude indeed always interferes constructively with the SM amplitude, indicating an enhancement of the branching ratio.

\begin{table}[th]
\begin{tabular}{cccccc}
\hline\hline
Observable & ~~~~~~$\mu_{ZZ}$~~~~~~ & ~~~~~~$\mu_{WW}$~~~~~~ & ~~~~~~$\mu_{\gamma\gamma}$~~~~~~
& ~~~~~~$\mu_{bb}$~~~~~~ & ~~~~~~$\mu_{\tau\tau}$~~~~~~ \\
\hline
ATLAS\cite{ATLAS2015} & $1.44^{+0.40}_{-0.33}$ & $1.09^{+0.23}_{-0.21}$ & $1.17 \pm 0.27$
& $0.52 \pm 0.40$ & $1.43^{+0.43}_{-0.37}$ \\
CMS~\cite{Khachatryan:2014jba} & $1.00\pm0.29$ & $0.83\pm0.21$ & $1.12\pm0.24$ & $0.84\pm0.44$
& $0.91\pm0.28$ \\
\hline\hline
\end{tabular}
\caption{Signal strengths of various modes, indicated by the subscript in the first column, as measured at the LHC.}
\label{tab:sigs}
\end{table}

With the dimension-6 operator Wilson coefficients we can fit to the electroweak precision data and the Higgs data, to get constraints for the Lagrangian parameter of VL mass $M$ and Yukawa couplings $y_u$ and $y_d$\footnote{For model independent constraints, see~\cite{Pomarol:2013zra,Ellis:2014dva}}. We do for both existing data and future expected measurements. For current measurements we refer to the $U=0$ oblique parameter measurements of~\cite{Baak:2014ora} and the ATLAS (CMS) Higgs data with $4.5~(5.1)~\text{fb}^{-1}$ integrated luminosity at $\sqrt{s}=7$~TeV and $20.3~(19.7)~\text{fb}^{-1}$ integrated luminosity at $\sqrt{s}=8$~TeV~\cite{ATLAS2015,Khachatryan:2014jba}, as listed in Table~\ref{tab:sigs}. We do not include tri-gauge boson precision measurements in our fitting. For future expected sensitivities, we take the most aggressive oblique parameter measurements expected from the TeraZ experiment~\cite{TeraZ} and the projected Higgs data from Table~4 of~\cite{Gomez-Ceballos:2013zzn}.

\begin{figure}[t!]
\centering
\includegraphics[height=3in]{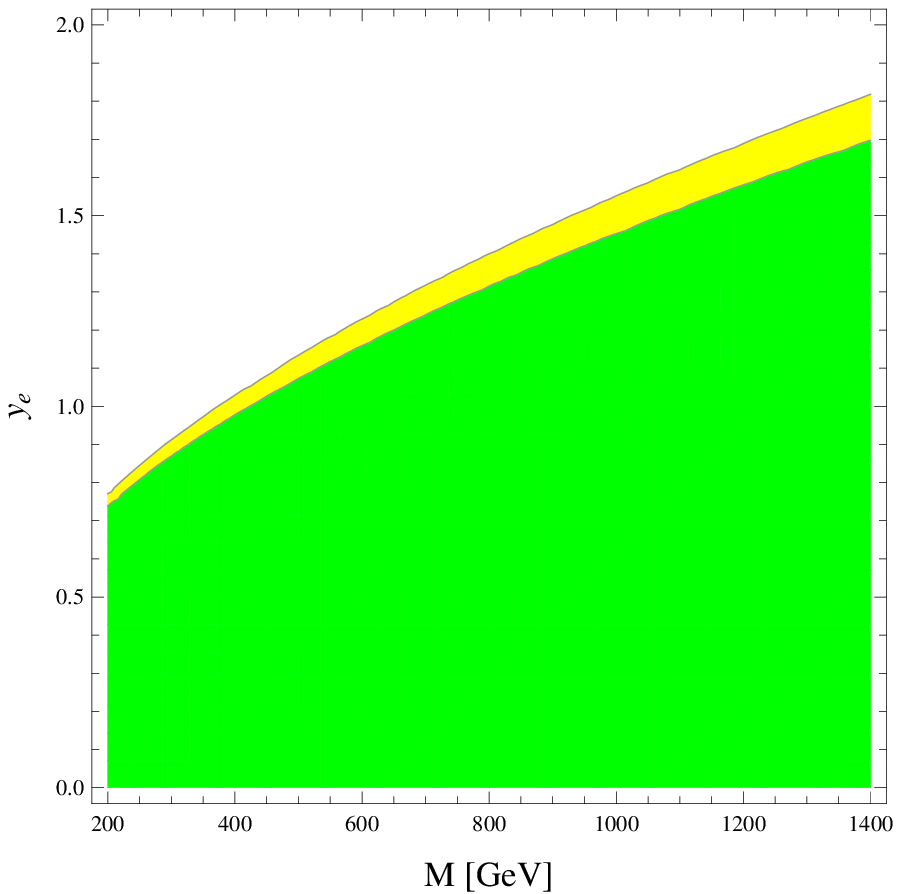}
\includegraphics[height=3in]{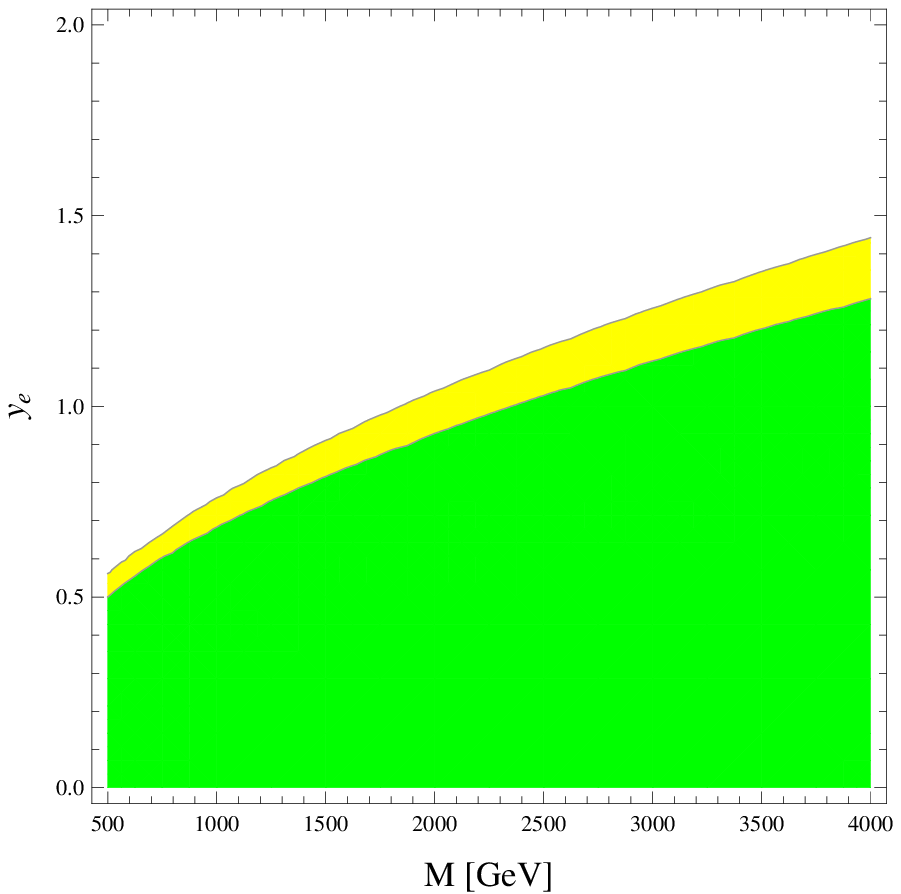}
\caption{Constraints on the $M$-$y_e$ parameter space of the VL lepton ($m=1,Y=-\frac{1}{2}$) model from a global fit to both the electroweak oblique corrections and the Higgs data. We have set the neutrino Yukawa $y_\nu=0$. The left plot uses the current LEP2 oblique parameters and the current ATLAS and CMS Higgs data. The right plot uses the most aggressive TeraZ result for the oblique parameters and the expected FCC-ee Higgs measurement constraints. Green and yellow regions are allowed regions at $1\sigma$ and $2\sigma$ level respectively.}
\label{fig:L}
\end{figure}

\begin{figure}[t!]
\centering
\includegraphics[height=3in]{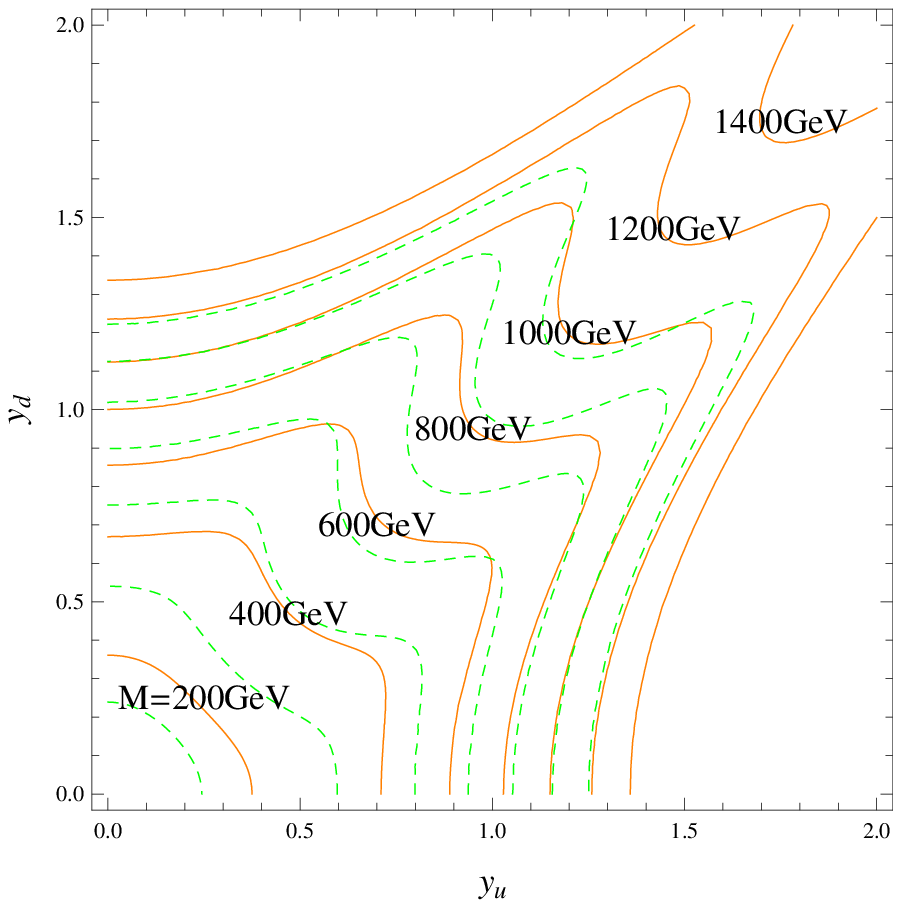}
\includegraphics[height=3in]{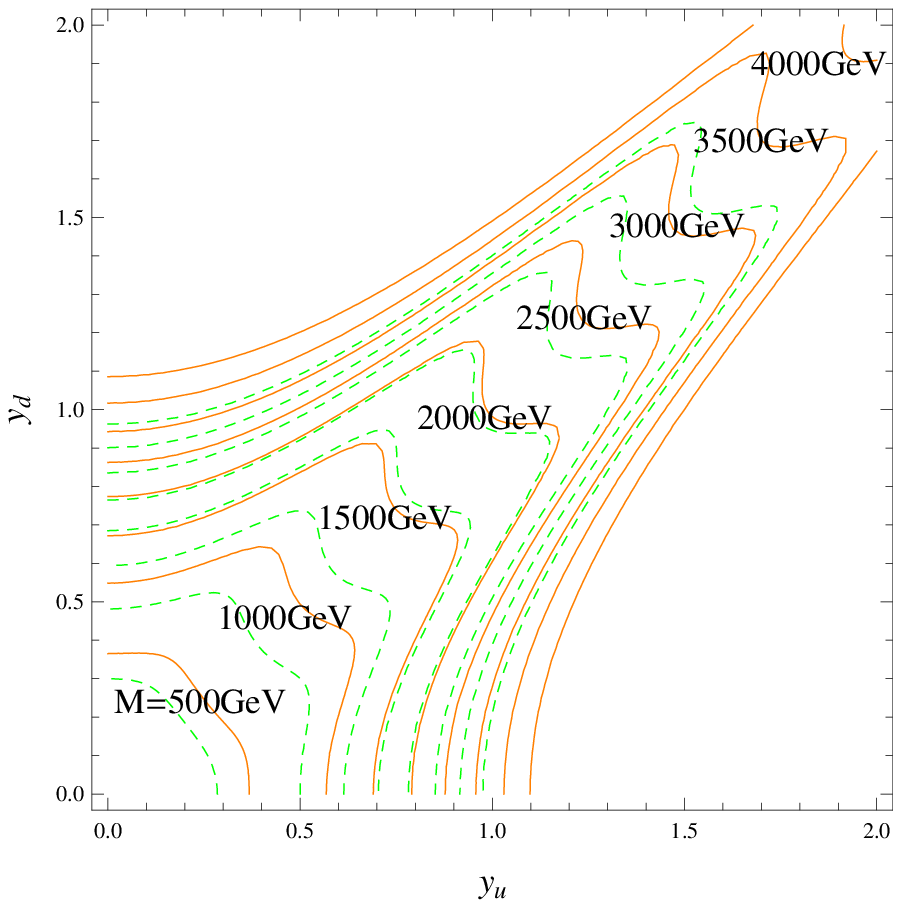}
\caption{Constraints on the $y_u$-$y_d$ parameter space of the VL quark ($m=3,Y=\frac{1}{6}$) model from a global fit to both the electroweak oblique corrections and the Higgs data. The left plot uses the current LEP2 oblique parameters and the current ATLAS and CMS Higgs data. The right plot uses the most aggressive TeraZ result for the oblique parameters and the expected FCC-ee Higgs measurement constraints. Green dashed and orange curves are contours at $1\sigma$ and $2\sigma$ levels for different choices of $M$ respectively. In the left plot we show contours for $M=200,400,600,800,1000,1200,1400$~GeV and in the right plot we show contours for $M=500,1000,1500,2000,2500,3000,3500,4000$~GeV.}
\label{fig:Q}
\end{figure}

In Fig.~\ref{fig:L} we show the numerical constraints for a VL lepton sector, in which we set the neutrino Yukawa coupling $y_\nu=0$. In addition to the oblique $S$ and $T$ parameter constraints, the Higgs diphoton branching ratio calculations in~\cite{Joglekar:2012vc,ArkaniHamed:2012kq} now is turned over to a constraint as well. Mostly probing the $y_e/M$ direction or so, our results are complementary to the direct VL lepton search, which has a mass bound of about $176$~GeV~\cite{Aad:2015dha}.

In Fig.~\ref{fig:Q} we show the numerical constraints for a VL quark sector. For small VL mass the constraint curves scale more like $y_u^2+y_d^2$, and indeed the $\chi^2$ is dominated by $\mathcal{O}_{GG}$ contribution to the gluon fusion in the current LHC data. As the VL mass increase the constraint is weaker in the direction of $y_u\simeq y_d$, which is an indication that the oblique correction $T$ parameter is becoming more and more dominant in that case. Since every Higgs field is accompanied by a Yukawa coupling in the CDE, the Wilson coefficient of a four Higgs operator such as $\mathcal{O}_T$ will scale quartically with the Yukawa coupling, being very sensitive except for the $y_u\simeq y_d$ direction, in case that the Yukawa couplings are large. The direct search constraints for the VL quark masses are $700\sim900$~GeV~\cite{Aad:2015kqa,Aad:2015mba,Khachatryan:2015axa,Khachatryan:2015gza}.

\section{Conclusion and Discussion\label{sec:dis}}

The SM EFT is a generalization of the precision measurements observables such as the electroweak oblique corrections and the Higgs precision measurements. At one-loop level the CDE of the CW potential provides a systematic tool of matching the UV theory to the SM EFT, all the Wilson coefficients of the dimension-6 operators are calculated if we enumerate all the contributing patterns in the CDE.

The merit of the CDE of CW potential approach includes:
\begin{itemize}
\item Complete: Up to pure Higgs and gauge field operators, all the Wilson coefficients of the dimension-6 operators are calculated as long as there is a contribution. On the other hand, it can match UV theory to redundant operators, then one can perform field redefinition or use equation of motion to remove redundant operators later on, as one wish.
\item Analytical: The CDE do not need to know the mass eigenstates and rotation matrices from interaction eigenstate basis to mass eigenstate basis. The complication of mixing is equivalent to the non-commutative nature of the matrices of $\delta\tilde{V}''+\tilde{G}+\tilde{\Gamma}_1+\tilde{\Gamma}_2$ with each other and with matrices of $M^{-2}$. And the final results of the Wilson coefficients in forms of some power of couplings divided by large mass squares, are equivalent to expanding the interaction-to-mass-eigenstate mixing matrix to power of $\mathcal{O}(M^{-2})$.
\item Model Independent Formulism: The input is just the BSM particles coupling matrices to the Higgs and SM gauge bosons, and all the following calculation are identical for different model.
\end{itemize}
Different from the Feynman diagram approach (together with the interaction-to-mass-eigenstate mixing), the CDE always gives only the leading terms in expansion of large mass square suppression. The expansion is also precise in case that the BSM coupling is small. Even if the current experiments are still exploring a region in which the eligibility of the CDE is questionable, as the experiments goes to higher and higher precisions and energies, if no deviation from the SM is detected, the eligibility of the CDE will become better and better.

So far the CDE is only calculated for the case that all the large masses are equal, and the matrix of the large scale square is proportional to an identity matrix so that commute with $\delta\tilde{V}''+\tilde{G}+\tilde{\Gamma}_1+\tilde{\Gamma}_2$. It is not hard to generalize it to the case that the large masses are nondegenerate~\cite{Drozd:2015kva}, if we keep track carefully of the order of the generally non-commutative matrices.

\acknowledgments

This research was supported in part by the World Premier International Research Center Initiative, Ministry of Education, Culture, Sports, Science and Technology, Japan.

\appendix*
\section{List of Contributing Patterns to Each Operator\label{sec:col}}

Here we collect all the combinations in the CDE of Eq.~(\ref{eq:CDEF}) contributing to the dimension-6 operators in Table~\ref{tab:Op}. As mentioned before the contributing terms can be sorted as $\gamma$ matrices irrelevant ones and $\gamma$ matrices relevant ones. The former group is common in both bosonic theory and fermionic theory, with a correspondence of fermionic $\{M,\delta V''_F\}$ and $(\delta V''_F)^2$ terms to the bosonic counterparts of dimension-1 and dimension-2 parts of $\delta V''$. In the following with a little bit misuse of terminology we will call the former $\gamma$ matrices irrelevant group as ``bosonic'', and the latter $\gamma$ matrices relevant terms as ``fermionic''. If the theory is indeed bosonic, then one only need to count the pure ``bosonic'' contributions for each operator.

The $\mathcal{O}_6$ is free of derivatives. It is given by all combination of the $(\delta V''_F)^2$ and $\{M,\delta V''_F\}$ which make up a total dimension of 6. It can be given by any permutation of three $(\delta V''_F)^2$s, or two $(\delta V''_F)^2$s together with two $\{M,\delta V''_F\}$s, or one $(\delta V''_F)^2$ together with four $\{M,\delta V''_F\}$s, or six $\{M,\delta V''_F\}$s.

The $\mathcal{O}_H$, $\mathcal{O}_T$ and $\mathcal{O}_R$ form a complete basis for all four Higgs two derivatives terms. In a bosonic theory such terms in the CDE come from a permutation of two $(\delta V''_F)^2$s, or one $(\delta V''_F)^2$ together with two $\{M,\delta V''_F\}$s, or four $\{M,\delta V''_F\}$s. A total of two covariant derivatives $-iD_\mu\frac{\partial}{\partial p}^\mu$ should appear at all possible positions; if both act on the same $(\delta V''_F)^2$ or $\{M,\delta V''_F\}$, then an integration by part should be used and one $D_\mu$ is moved to all the other factors. For fermionic theory in addition to the bosonic contribution, there are contributions from permutations of two $i\slashed{D}\delta V''_F$s (first term of Eq.~(\ref{eq:Gamma1})) together with one $(\delta V''_F)^2$, and two $i\slashed{D}\delta V''_F$s together with two $\{M,\delta V''_F\}$s.

Since we keep all the redundant operators and do not use relations such as Eq.~(\ref{eq:HWB}) for projection, $\mathcal{O}_{GG}$, $\mathcal{O}_{WW}$, $\mathcal{O}_{BB}$ and $\mathcal{O}_{WB}$ only receive contributions from the CDE combinations where two Higgs and two gauge field strengths directly enter. In a bosonic theory such terms are permutations of a $(\delta V''_F)^2$ together with a $\frac{1}{4}g^2t^at^bF^a_{\nu\mu}F^{b\rho\mu}\frac{\partial}{\partial p}^\nu\frac{\partial}{\partial p}_\rho$ (fourth term of Eq.~(\ref{eq:expG})), or two $\{M,\delta V''_F\}$s together with a $\frac{1}{4}g^2t^at^bF^a_{\nu\mu}F^{b\rho\mu}\frac{\partial}{\partial p}^\nu\frac{\partial}{\partial p}_\rho$. In a fermionic theory additional terms are permutations of one $(\delta V''_F)^2$ together with two $-\frac{i}{4}g[\gamma^\mu,\gamma^\nu]F_{\mu\nu}^at^a$s (first term of Eq.~(\ref{eq:Gamma2})), or two $\{M,\delta V''_F\}$s together with two $-\frac{i}{4}g[\gamma^\mu,\gamma^\nu]F_{\mu\nu}^at^a$s, or two $\frac{i}{2}g\gamma^\mu[\delta V''_F,\thinspace F^a_{\nu\mu}t^a]\frac{\partial}{\partial p}^\nu$s (first term of second line of Eq.~(\ref{eq:Gamma1})).

The operators $\mathcal{O}_W$, $\mathcal{O}_B$ and $\mathcal{O}_{HW}$, $\mathcal{O}_{HB}$ are all two Higgs one gauge field strength and two derivatives operators. The integration by parts trick which moves the covariant derivative $D$ to other positions can mix the two groups and therefore cannot be used any more. The contributing combinations to $\mathcal{O}_W$ and $\mathcal{O}_B$ in a bosonic theory are just permutations of one $\frac{4}{3!}gp^\mu t^aD_\rho F_{\nu\mu}^a{\textstyle\frac{\partial}{\partial p}^\rho\frac{\partial}{\partial p}^\nu}+\frac{2}{3!}gt^aD^\mu F_{\nu\mu}^a{\textstyle\frac{\partial}{\partial p}^\nu}$ (first and second terms of Eq.~(\ref{eq:expG})) together with one $-iD_\mu\{M,\delta V''_F\}\frac{\partial}{\partial p}^\mu$ and one $\{M,\delta V''_F\}$. In fermionic theory additional contribution comes from permutations of a $\frac{2}{3!}g\gamma^\mu[\delta V''_F,\thinspace D_\rho F^a_{\nu\mu}t^a]{\textstyle\frac{\partial}{\partial p}^\nu\frac{\partial}{\partial p}^\rho}$ (third term of second line of Eq.~(\ref{eq:Gamma1})) together with a $i\slashed{D}\delta V''_F$, there are also contributions from the fermionic two Higgs four derivatives terms.

The combinations contributing to $\mathcal{O}_{HW}$ and $\mathcal{O}_{HB}$ are, in bosonic theory a permutation of one $igp^\mu t^aF_{\nu\mu}^a{\textstyle\frac{\partial}{\partial p}^\nu}$ (first term in Eq.~(\ref{eq:expG})) with two $-iD_\mu\{M,\delta V''_F\}\frac{\partial}{\partial p}^\mu$. There are a lot of fermionic contributions, from permutations of 
one $-\frac{i}{4}g[\gamma^\mu,\gamma^\nu]F_{\mu\nu}^at^a$ together with two $i\slashed{D}\delta V''_F$s, one $\frac{1}{2}g\gamma^\mu[D_\rho\delta V''_F,\thinspace F^a_{\nu\mu}t^a]\frac{\partial}{\partial p}^\nu\frac{\partial}{\partial p}^\rho$ together with one $i\slashed{D}\delta V''_F$, and contributions from the fermionic two Higgs four derivatives terms.

The bosonic contribution to two Higgs four derivatives operator $\mathcal{O}_D$ is a permutation of two $\{M,\delta V''_F\}$ together with four covariant derivatives acting on all possible positions, and the fermionic contribution should be a permutation of two $\frac{1}{2}(\slashed{D}D_\mu+D_\mu\slashed{D})\delta V''_F{\textstyle\frac{\partial}{\partial p}^\mu}$s (second term of the first line of Eq.~(\ref{eq:Gamma1})), or one $i\slashed{D}\delta V''_F$ together with one $-\frac{i}{6}(\slashed{D}D_\mu D_\nu+D_\mu\slashed{D}D_\nu+D_\mu D_\nu\slashed{D})\delta V''_F{\textstyle\frac{\partial}{\partial p}^\mu\frac{\partial}{\partial p}^\nu}$ (third term of the first line of Eq.~(\ref{eq:Gamma1})). The integration by parts trick can be used to always move two derivatives to one Higgs and the other two to the other, in accord with the form of $\mathcal{O}_D$. Note that the four covariant derivatives are eventually contracted by two metric $g^{\mu\nu}$s, and the desired $\mathcal{O}_D$ corresponds to a specific order of the two $D_\mu$s and two $D_\nu$s. In order to get to the desired order one need to commute $D_\mu$ and $D_\nu$. In bosonic theory all the gauge field strength from the commutator of $[D_\mu,\thinspace D_\nu]=-igF_{\mu\nu}^at^a$ has been collected and resummed by the expansion of $\tilde{G}$, so the commutator should not be counted. But in fermionic theory the resummation of gauge field strength terms has not been done in the $\gamma$ matrix related term, not all of them are collected by the $\tilde{G}$ or $\tilde{\Gamma}_2$. The $[D_\mu,\thinspace D_\nu]=igF_{\mu\nu}^at^a$ in switching orders among the four derivatives should be counted, which makes additional contribution to the $\mathcal{O}_W$, $\mathcal{O}_B$ and $\mathcal{O}_{HW}$, $\mathcal{O}_{HB}$. In particular, organizing terms to the form of two derivatives acting on one Higgs and two on the other, there can be three different orders: directly $(D_\mu D^\mu H^\dag)(D_\nu D^\nu H)=\mathcal{O}_D$, as well as $(D_\mu D_\nu H^\dag)(D^\mu D^\nu H)$ and $(D_\mu D_\nu H^\dag)(D^\nu D^\mu H)$. $(D_\mu D_\nu H^\dag)(D^\mu D^\nu H)$ will have a contribution to $\mathcal{O}_W$, $\mathcal{O}_B$ as well as $\mathcal{O}_{HW}$, $\mathcal{O}_{HB}$, and $(D_\mu D_\nu H^\dag)(D^\nu D^\mu H)$ will have a contribution only to $\mathcal{O}_{HW}$, $\mathcal{O}_{HB}$. To evaluate the contribution a transformation based on Jacobi identity is found useful.


\end{document}